\documentclass[aps,pre,preprint,superscriptaddress]{revtex4}

\usepackage{graphicx}
\usepackage{dcolumn}
\usepackage{bm}

\begin{document}


\title{RANDOM FIELD CRITICAL SCALING IN A MODEL OF DIPOLAR GLASS
AND RELAXOR FERROELECTRIC BEHAVIOR}


\author{Ronald Fisch}
\affiliation{382 Willowbrook DR, North Brunswick, NJ 08902}


\date{\today}

\begin{abstract}
Heat bath Monte Carlo simulations have been used to study a 12-state
discretized Heisenberg model with a new type of random field on
simple cubic lattices of size $128 \times 128 \times 128$.  The 12 states
correspond to the [110] directions of a cube.  The model has the standard
nonrandom two-spin exchange term with coupling energy $J$ and a random field
which consists of adding an energy $h_R$ to two of the 12 spin states, chosen
randomly and independently at each site.  We report on the case $h_R / J = 3$,
which has a sharp phase transition at about $T_c / J = 1.40625$.  Below $T_c$,
the model has long-range ferroelectric order oriented along one of the eight
[111] directions.  At $T_c$, the behavior of the peak in the structure factor,
$S ({\bf k} )$, at small $|{\bf k}|$ is a straight line on a log-log plot,
which gives the result $\bar{\eta} = 1.214 \pm 0.014$.  The onset of
orientational order below $T_c$ is very rapid for this value of $h_R$.  There
are peaks in the specific heat and longitudinal susceptibility at $T_c$.
Below $T_c$ there is a strong correction to ordinary scaling, which is probably
caused by the cubic anisotropy, which is a dangerous irrelevant variable.

\end{abstract}


\maketitle


\section{INTRODUCTION}

The ${\bf O_{12}}$ model\cite{Fis98,Fis98b} is a model of discretized Heisenberg
spins, {\it i.e.} ${\bf O_{12}}$ is a discrete subgroup of ${\bf O(3)}$.  We will
put this model on a simple cubic lattice of size $L \times L \times L$, with
periodic boundary conditions.  It is now almost 25 years since the original Monte
Carlo study\cite{Fis98} of the three-dimensional (3D) random field $\bf O_{12}$
model.  The numerical results presented there, which used $L \le 64$, are crude by
current standards, but the reasons given for why this model is worthy of study are
just as valid now as they were then.  This model is designed to represent a
ferroelectric dipolar glass.\cite{HKL90,BR92}  It is also believed to be relevant\cite{HV76}
for ferroelectric phases in strongly disordered cubic perovskite alloys, which are
often referred to as relaxor ferroelectrics.\cite{SI54,BD83,WKG92,Kle93,PTB87,PB99}
From this point of view, the essential element for relaxor ferroelectric behavior
is the alloy disorder, rather than a particular chemistry.  A cubic perovskite
which has a supercell ordering of a few perovskite unit cells in size should not be
considered a relaxor ferroelectric, regardless of its chemical composition.

The work reported here uses modern computing power and a somewhat improved
Monte Carlo algorithm to obtain results for lattices of size $L = 128$, which
was not possible in the earlier work.  We also make a more sophisticated choice
of the random field distribution to obtain new results which should be of great
interest to many people.  We feel that the new results presented here demonstrate
that the hope of Halperin and Varma\cite{HV76}, that this type of model can provide
new insights into the nature of perovskite ferroelectric materials, is finally
being realized.

Our Hamiltonian starts with the classical Heisenberg exchange form for a ferromagnet:
\begin{equation}
  H_{ex} ~=~ - J \sum_{\langle ij \rangle} {\bf S}_{i} \cdot {\bf S}_{j}.
\end{equation}
Each spin ${\bf S}_{i}$ is a three-component dynamical variable which has
twelve allowed states, which are unit vectors pointing the [110] directions.
The $\langle ij \rangle$ indicates a sum over nearest neighbors of the simple
cubic lattice.  The symbol $H$ denotes the uniform external field.  It does
not appear explicitly in our Hamiltonian, because we have set $H$ to zero.
There is no loss of generality in setting $J = 1$.  We will also set Boltzmann's
constant to 1.  This merely establishes the units of energy and temperature.
The restriction of the spin variables to ${\bf O_{12}}$ builds a
temperature-dependent cubic anisotropy into the model.  A useful review of the
effects of such a cubic anisotropy has been given by Pfeuty and Toulouse.\cite{PT77}

We add to $H_{ex}$ a random field which, at each site, is the sum of two Potts
field terms:
\begin{equation}
  H_{RP2} ~=~ h_{R} \sum_{i} (\delta_{{\bf S}_{i} , {\bf \hat{u}}_{i}} +
\delta_{{\bf S}_{i} , {\bf \hat{v}}_{i}})   \, .
\end{equation}
Each ${\bf \hat{u}}_{i}$ and ${\bf \hat{v}}_{i}$ is an independent quenched random
variable. which assumes one of the twelve [110] allowed states with equal
probability.  Since ${\bf \hat{u}}_{i}$ is allowed to be equal to ${\bf \hat{v}}_{i}$,
the random field at each site has 78 possible types.  The reader should note that,
for an $XY$ model, adding a second Potts field would generate nothing new, due to the
Abelian nature of the ${\bf O(2)}$ group.  However, in the current case, adding the
second Potts field can substantially reduce the corrections to finite size scaling.

The celebrated Imry-Ma construction,\cite{IM75} argues that random fields
built from groups with continuous symmetries cannot have conventional long-range
order when the number of spatial dimensions is less than or equal to four.
The Imry-Ma argument implicitly assumes that the spin group is Abelian.  What is
assumed is that the coupling of the random field to the spins is purely of vector
form. However, at a renormalization group fixed point, a nonabelian random vector
field has the ability to generate higher order tensor couplings to the effective
spins.  Whether or not such higher order couplings are relevant at the critical
fixed point must be studied explicitly, on a case-by-case basis.

The results we find here demonstrate that the Imry-Ma instability does not
necessarily exist for random fields applied to a Heisenberg model.  For a
nonabelian spin group the Halperin-Saslow effect\cite{HS77} changes the
long-wavelength dispersion from quadratic to linear.  This can cure the Imry-Ma
instability.  This mechanism is also closely related to the ordering in the 3D
Heisenberg spin glass.\cite{FMPTY09}  From the hydrodynamic viewpoint, a spin
glass and a random field model for the same type of spins should have the same
lower critical dimension.  However, the Imry-Ma argument can only be applied to
systems which do not have a Kramers degeneracy. The problem with the Imry-Ma
argument for the Heisenberg model is that for nonabelian spins the sum of two
noncollinear vectors does not transform like a vector.

The reader may be tempted to object that the analysis of Halperin and Saslow
should not be applicable in the presence of the cubic anisotropy.  However, what
we are really interested in here is the issue of the stability of the random-field
Heisenberg critical point.  It is believed that in 3D the pure Heisenberg critical
point is stable against the presence of cubic anisotropy favoring the [111]
directions.\cite{PT77}  For the ${\bf O}_{12}$ model, this was verified numerically
by the author.\cite{Fis98}  Here we will demonstrate numerically that the necessary
condition for the 3D random-field Heisenberg critical point to be stable against
[111] cubic anisotropy is also satisfied.  This type of cubic anisotropy is often
referred to as a "dangerous irrelevant variable".

In the work presented here, we will set $h_R = 3$.  Study of the model
for other values of $h_R$ is under way.  Note that having $h_R$ positive
means that ${\bf S}_{i}$ has an increased energy if it points in the direction
of ${\bf \hat{u}}_{i}$ or ${\bf \hat{v}}_{i}$.  As shown by density functional
calculations,\cite{CGMR02} and studied previously in the author's work on a
four-state model,\cite{Fis03} a positive $h_R$ is an appropriate choice for a
model of a relaxor ferroelectric.  The author believes that a negative value of
$h_R$ should be used for modeling a dipolar glass.

It is likely, however, that critical exponents will not depend on whether $h_R$
is positive or negative, or even whether the values of $h_R$ are equal for
${\bf \hat{u}}_{i}$ and ${\bf \hat{v}}_{i}$.  Of course, we do not know yet
whether this is true.  In any case, the size of corrections to scaling are not
universal.  The author expects that corrections to scaling will be smallest when
$h_R$ is the same for both ${\bf \hat{u}}_{i}$ and ${\bf \hat{v}}_{i}$, because
that is when the Halperin-Saslow effect is the strongest.  The reader should note
that, with this type of a random field, the model does not become trivial in the
limit $h_R \to \infty$.

The range of distributions of the random fields for which this effect will be
found remains unclear.  Note that the pure Heisenberg critical point is not stable
against the introduction of cubic anisotropy which favors the [100] directions.
In that case, the phase transition becomes first order.\cite{PT77}  However, a
first order phase transition with a latent heat is not allowed in a model with
quenched disorder.\cite{AW89,AW90}

\section{NUMERICAL PROCEDURES}

If $h_R$ is chosen to be an integer, then the energy of any state of the
model is an integer.  Then it becomes straightforward to design a heat-bath
Monte Carlo computer algorithm to study it which uses integer arithmetic
for the energies, and a look-up table to calculate Boltzmann factors.
This procedure substantially improves the performance of the computer
program, and was used for all the calculations reported here.  (The program
currently has the ability to use half-integer values of $h_R$.) Lattices
with periodic boundary conditions were used throughout.

Three different linear congruential random number generators are used.
The first chooses the ${\bf \hat{u}}_{i}$, the second chooses the
${\bf \hat{v}}_{i}$ and the third is used in the Monte Carlo routine which
flips the spins.  The generator used for the spin flips, which needs very
long strings of random numbers, was the Intel library program $random\_number$.
In principle, Intel $random\_number$ can be used for multicore parallel
processing.  However, our program is so efficient in single-core mode that no
speedup was seen when the program was run in parallel mode.

The code was checked by correctly reproducing the known results\cite{Fis98}
of the $h_R = 0$ case, and extending them to $L = 128$.  For $h_R = 3$,
32 samples of size $L = 128$ were studied.  The same set of 32 samples was
used for all values of $T$, so it is meaningful to talk about heating and
cooling of a sample.  Each sample was initially in a random state, and then
cooled slowly, starting from $T$ = 1.5625, and ending at $T$ = 1.3125.
Both cooling and heating were done in temperature steps of 0.015625, with times
of 20,480 Monte Carlo steps per spin (MCS) at each stage.  (The reader should note
that the temperatures we use in the computer code are simple binary fractions,
even though they may appear to be unnatural when written in decimal notation.)

For the pure ${\bf O_{12}}$ system, the Heisenberg critical temperature\cite{Fis98}
is $T_c$ = 1.453.  For $h_R = 3$, the spin correlations are only short-ranged
above this temperature, so no extended data runs were made in this region of $T$.
For 30 of the 32 samples studied, the system was strongly oriented along one
of the [111] directions by the time it had been cooled to $T$ = 1.3125.  The
other two samples had become trapped in metastable states.  These two samples
were then initialized in [110] states which had a positive overlap of approximately
0.5 with their metastable states.  They were run at $T$ = 1.25,  where they
were easily able to relax to low energy [111] states.  These states were
then warmed slowly to $T$ = 1.4375.  At $T$ = 1.34375, these [111] states were
seen to have significantly lower energies than the metastable states found by
cooling for those two samples.

The fact that it was not difficult to find a stable state oriented along [111]
at $T$ = 1.25 means that, for $h_R = 3$, an $L = 128$ sample does not have many
metastable states.  However, it is expected that for large enough values of $L$
it should be true that in the [111] ferroelectric phase there would be a
metastable state corresponding to each of the [111] directions.  Whether or not
this will continue to be true for larger values of $h_R$ is not yet known.
In the proposed power-law correlated phase,\cite{Fis98} it would no longer be
possible to assign low-energy metastable states to particular [111] directions.

Two trial runs made on $L = 64$ lattices indicated the presence of a [111] to
[110] phase transition at roughly $T$ = 0.875 for $h_R = 3$.  However, a
detailed study of this second transition was not undertaken.

Extended runs for data collection using the hot initial condition were made at
$T$ = 1.4375, 1.40625, 1.375 and 1.34375.  For the cold initial condition, data
collection runs were made at $T$ = 1.34375, 1.375, 1.390625 and 1.40625.  As we
shall see, relaxation of the hot and cold initial conditions gave essentially
indistinguishable results at $T$ = 1.40625.

A data collection run for each sample consisted of a relaxation stage and a
data stage.  For the three lower values of $T$, a relaxation stage was a run
of length 122,880 MCS.  In most cases this was sufficient to bring the sample
to an apparent local minimum in the phase space.  This was followed by a data
collection stage of the same length.  If further relaxation was observed during
the data collection stage, it was reclassified as a relaxation stage.  Then a
new data collection stage was run.  The energy and magnetization of each sample
were recorded every 20 MCS.  A spin state of the sample was saved after each
20,480 MCS.  Thus there were six spin states saved from the data collection
stage for each run.  These six spin states were Fourier transformed and averaged
to obtain the structure factor for each sample.  At $T$ = 1.4375, a relaxation
stage of only 20,480 MCS was judged to be sufficient.

\section{THERMODYNAMIC FUNCTIONS}

The thermodynamic data calculated from our Monte Carlo data on the 32 $L = 128$
samples are shown in Table I.

For a random field model, unlike a random bond model, the average value of
the local spin, $\langle {{\bf S}_i} \rangle$, is not zero even in the
high-temperature phase.  The angle brackets here denote a thermal average.
Thus the longitudinal part of the susceptibility,
$\chi_{||}$, is given by
\begin{equation}
  T \chi_{||} ({\bf k}) ~=~ 1 - |{\bf M}|^2 ~+~ L^{-3} \sum_{ i \ne j } \cos (
  {\bf k}  \cdot {\bf r}_{ij}) ({\bf S}_{i} \cdot {\bf S}_{j} ~-~ Q_{ij} )  \,   ,
\end{equation}
For Heisenberg spins,
\begin{equation}
  Q_{ij} ~=~ \langle {{\bf S}_i} \rangle \cdot \langle {{\bf S}_j} \rangle  \,  ,
\end{equation}
and
\begin{equation}
  |{\bf M}|^2 ~=~ L^{-3} \sum_{i} Q_{ii}
      ~=~ L^{-3} \sum_{i} [ \langle {\bf S}_{i} \rangle \cdot \langle {\bf S}_{i} \rangle ]_t \,  .
\end{equation}
where the square brackets $[ ... ]_t$ indicate a time average.   $Q_{ij}$ must be
included for all $T$.  Note that the $\chi_{||}$ we define here is not exactly what
would be called the longitudinal susceptibility in a nonrandom system.  The distinction
between longitudinal modes and transverse modes is not completely well-defined in a system
which has a local order parameter that is not full aligned with the sample-averaged order
parameter.  In any case, our system has no soft modes below $T_c$ in the ferroelectric
phase, due to the cubic anisotropy.

The specific heat, $c_H$, may be calculated by taking the finite differences
$\Delta E / \Delta T$, where $E$ is the energy per spin.  Alternatively, it
may be calculated as the variance of $E$ divided by $T^2$.  The second method
was used for the $c_H$ numbers shown in Table I.

The data in Table I show that there are peaks in $\chi_{||}$ and $c_H$ at $T_c$.
However, accurate estimates of the values of the critical exponents $\alpha$
and $\gamma$ would require collecting a lot more data at temperatures close
to $T_c$.  As we shall see shortly, our data below $T_c$ do not take the form of
the usual critical scaling behavior.  This may be partly a reflection of the fact
that this type of model is expected to show replica-symmetry breaking\cite{MY92}
RSB) below $T_c$.  It has been understood for a long time that, in a mean-field
approximation, RSB is a type of ergodicity breaking.\cite{SZ82}  We are very
far from mean-field theory here, however.

Define $M_x (t)$, $M_y (t)$ and $M_z (t)$ to be the averages over the lattice at
time $t$ of the components of ${\bf S}_i$, in the usual way.  Then the cubic
orientational order parameter (COO) is measured by calculating the quantity
\begin{equation}
  COO = 3 [ ({M_x}^2 {M_y}^2 + {M_x}^2 {M_z}^2 + {M_y}^2 {M_z}^2) / |{\bf M}|^4 ]_t   \, .
\end{equation}
The possible values of COO range from 0 when ${\bf M}$ points in a [100] direction
to 1 when ${\bf M}$ points along a [111] direction.  Note that if all of the spins
were fully aligned along one of the [110] directions, the value of COO would be $3/4$.
Table I shows that the sample average of COO is approximately 0.60 in the paraelectric
phase, and that it rapidly increases to 1 as $T$ is reduced below $T_c$.  This behavior
indicates that the [111] orientational ordering is irrelevant at the random-field
Heisenberg critical point, just as it is irrelevant at the pure Heisenberg critical
point.

\begin{quote}
\begin{flushleft}
Table I: Thermodynamic data for $128 \times 128 \times 128$
lattices at $h_R = 3$, for various $T$. (h) and (c) signify
data obtained relaxing from hotter and colder initial conditions,
respectively.  The one $\sigma$ statistical errors shown are due
to the sample-to-sample variations.
\begin{tabular}{|l|ccccc|}
\hline
$~~~~~~T$&$|{\bf M}|$&$\chi_{||}$&$E$&$c_{H}$&$COO$\\
\hline
1.34375(c)&0.4270$\pm$0.0024&25.9$\pm$3.8&-1.1955$\pm$0.0001&2.782$\pm$0.008&0.985$\pm$0.006\\
1.375~~~(c)&0.3108$\pm$0.0041&86.0$\pm$7.8&-1.1047$\pm$0.0002&2.997$\pm$0.012&0.869$\pm$0.029\\
1.390625(c)&0.2309$\pm$0.0051&146$\pm$26&-1.0569$\pm$0.0002&3.103$\pm$0.021&0.726$\pm$0.054\\
1.40625(c)&0.1253$\pm$0.0055&291$\pm$29&-1.0072$\pm$0.0002&3.185$\pm$0.018&0.616$\pm$0.041\\
1.4375~(h)&0.0238$\pm$0.0012&72.4$\pm$1.7&-0.9173$\pm$0.0001&2.472$\pm$0.009&0.599$\pm$0.012\\
\hline
\end{tabular}
\end{flushleft}
\end{quote}

The structure factor, $S ({\bf k}) = \langle |{\bf M}({\bf k})|^2 \rangle $,
for Heisenberg spins is
\begin{equation}
  S ({\bf k}) ~=~  L^{-3} \sum_{ i,j } \cos ( {\bf k} \cdot {\bf r}_{ij})
   {\bf S}_{i} \cdot {\bf S}_{j}  \,   ,
\end{equation}
where ${\bf r}_{ij}$ is the vector on the lattice which starts at site $i$
and ends at site $j$.  When there is a phase transition into a state with
long-range spin order, $S ( {\bf k} = 0 )$ has a stronger divergence than
$\chi ( {\bf k} = 0 )$ does.

Values of $S (|{\bf k}|)$ calculated by taking an angular average for each
sample and then averaging the results for the 32 $L = 128$ samples are shown
in Fig.~1.  In Fig.~2, the results for the 15 smallest non-zero values of
$|{\bf k}|$ at $T$ = 1.375, 1.390625 and 1.40625 are shown.  Note that the
data shown for the cold initial condition and the hot initial condition at
$T$ = 1.40625 are virtually indistinguishable.  At lower temperatures, two
of the hot initial condition samples became trapped in metastable states.

\section{DISCUSSION}

The RP2 random field distribution defined by Eqn.~2 is analogous to
a quantum random field built with spin 2 operators.  Since a single-site
operator term in a Hamiltonian of quantum ${\bf O(3)}$ spins which favors
the [111] directions is also a spin 2 operator, it seems natural that this
type of random field can give a direct transition from the paraelectric
phase into a ferroelectric phase having random field characteristics.  This
will only happen if $h_R$ is chosen appropriately.  The author expects that
when $h_R$ becomes very large, a phase with quasi-long-range order will be
seen.\cite{Fis98} It remains to demonstrate this.  If that does not happen,
then the next step would be to add a third random Potts field, {\it i.e.}
to study the RP3 model.

The fact that for the range of $|{\bf k}|$ shown in Fig.~2 the straight-line
fit to the data for $T$ = 1.40625 is essentially perfect is remarkable.  This
only a numerical accident, however, since we did not tune the temperature to
find this condition.  The fit of the hot initial condition data to a straight
line has a slope of $-2.788 \pm 0.014$, and the fit to the cold initial
condition data has a slope of $-2.784 \pm 0.015$.  Averaging these gives
\begin{equation}
  -(4 - \bar{\eta}) ~=~ -2.786 \pm 0.014  \,
\end{equation}
as the slope of $S(|{\bf k}|)$ on the log-log plot at the critical point
in the scaling region. We do not reduce our error estimate, because the
data for the different initial conditions on the same set of samples
cannot be considered to be statistically independent.  Thus
\begin{equation}
  \bar{\eta} ~=~ 1.214 \pm 0.014  \,
\end{equation}
which is a quite reasonable value for this quantity.  The value of $\bar{\eta}$
should not be sensitive to varying $T$ by a small amount away from $T$ = 1.40625.
This is shown by the fact that the data in Fig.~2 for $T$ = 1.390625 run parallel
to the data for $T$ = 1.40625 over a range of $|{\bf k}|$, and for smaller
$|{\bf k}|$ the slope of $S(|{\bf k}|)$ becomes more negative.  What this means
is that $\bar{\eta}$ is not a continuous function of $T$.  We are not seeing a
critical phase of the Kosterlitz-Thouless type.

This effect will eventually be cut off by the cubic anisotropy, which
stabilizes the ferroelectric order.  What this means is that there is a
strong correction to finite-size scaling of the ferroelectric order parameter
below $T_c$.  This effect was already seen in the author's earlier
work.\cite{Fis98}  This strong correction to scaling, which is dramatically
different from what is usually seen at a second order phase transition, is
likely what gives rise to the experimentally observed relaxor ferroelectric effect.
The author believes that this strong correction to scaling is a direct result
of the dangerous irrelevant variable, the cubic anisotropy.  This causes the
COO to vary rapidly as a function of $T$ just below $T_c$, as shown in Table~I.

For larger $|{\bf k}|$ the slope of $S$ is less negative, because the system
is then in the crossover region between the pure system fixed point and the
random-field fixed point.  A calculation for $h_R$ = 4, similar to the one
described here, is currently in progress.  The scaling region should then
extend to larger values of $|{\bf k}|$.  Incomplete results appear to show
that the value of $\bar{\eta}$ is smaller for $h_R$ = 4.  We are also
obtaining data at negative values of $h_R$.  The results for negative $h_R$
are qualitatively similar to the results for positive $h_R$.

We are not claiming to understand completely why this behavior occurs in our
fairly simple model.  We only claim that we have shown that this behavior, which
is similar to the observed central peak in $S(|{\bf k}|)$ seen in experiments,
does not require the inclusion of any additional feature, such as chemical
correlations, in the model.

\section{CONCLUSION}

In this work we have used Monte Carlo computer simulations to study
a model of discretized Heisenberg spins on simple cubic lattices in 3D,
with a carefully chosen type of random field.  We have found that,
for the random field distribution used here, our system shows a sharp
critical point and a transition into a phase with long-range order.
The phase transition shows characteristics expected of random field
behavior.  This example shows that the extreme difficulty in reaching
equilibrium which is usually associated with glassy behavior is not an
inevitable consequence of the presence of random fields.  Standard
renormalization group arguments imply that a similar transition ought
to exist in a 3D model of this type with continuous spins and random
fields which have an isotropic probability distribution.  The primary
reason why the existence of such behavior has been considered impossible
until now is that the nonabelian nature of the ${\bf O(3)}$ group, which
can give rise to relevant higher order tensor couplings to the random
field, was not properly taken into account.  We have also identified
the relaxor ferroelectric effect as arising from a strong correction
to scaling of the ferroelectric order parameter near $T_c$, which is
probably caused by the cubic anisotropy, which is present in this model
and also in the cubic perovskite relaxor ferroelectrics.  This strong
correction to scaling may not occur in a model with isotropic spins and
an isotropic distribution of random fields, or in an experimental system
which is not a crystalline alloy.

\begin{acknowledgments}

This work used the Extreme Science and Engineering Discovery Environment (XSEDE)
through allocation DMR180003.  Bridges Regular Memory at the Pittsburgh
Supercomputer Center was used for some preliminary code development, and the
new Bridges-2 Regular Memory machine was used to obtain the data discussed here.
The author thanks the staff of the PSC for their help.

\end{acknowledgments}


\newpage
\begin{figure}
\includegraphics[width=3.4in]{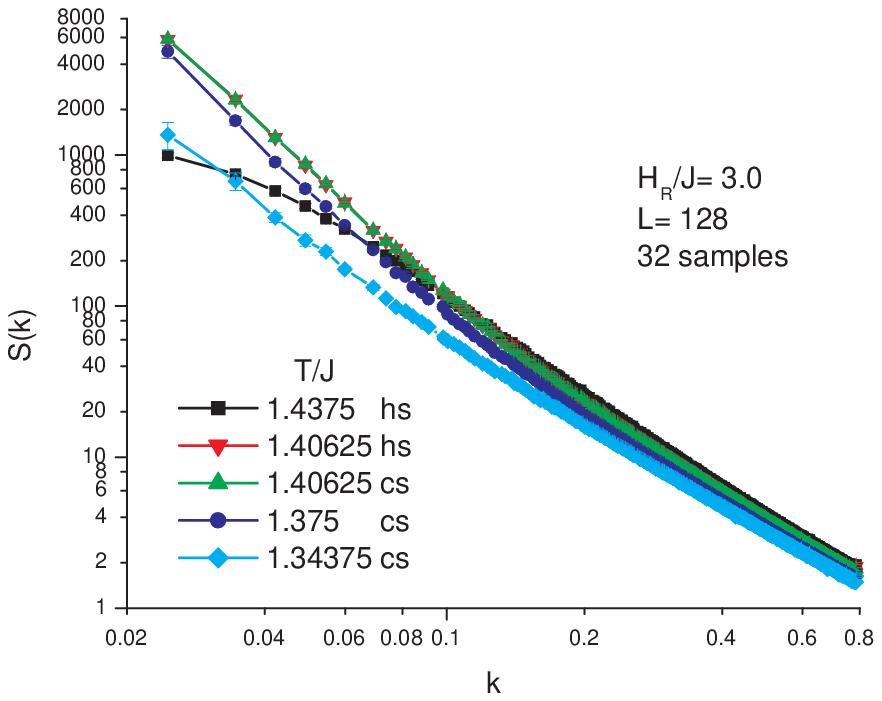}
\caption{\label{fig1} Angle-averaged magnetic structure factor, $S (|{\bf k}|)$,
at a sequence of temperatures for the RP2 ${\bf O}_{12}$ model with $h_R = 3$
on $128 \times 128 \times 128$ simple cubic lattices, log-log plot.  The points
show averaged data from 32 samples, at a series of temperatures. (h) and (c)
signify data obtained by relaxing from hotter and colder initial conditions,
respectively.}
\end{figure}

\begin{figure}
\includegraphics[width=3.4in]{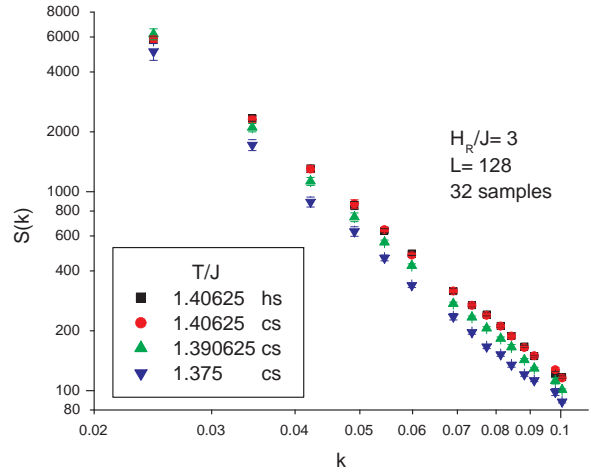}
\caption{\label{fig2} Angle-averaged magnetic structure factor, $S (|{\bf k}|)$,
at $T / J$ = 1.375, 1.390625 and 1.40625 for the RP2 ${\bf O}_{12}$ model
with $h_R = 3$ on $128 \times 128 \times 128$ simple cubic lattices, log-log plot.
The points show averaged data from 32 samples.  (h) and (c) signify data obtained
by relaxing from hotter and colder initial conditions, respectively.}
\end{figure}

\end{document}